\documentclass[fleqn,twoside]{article}
\usepackage{espcrc2}

\usepackage{graphicx}
\usepackage[figuresright]{rotating}

\newcommand{\AmS}{{\protect\the\textfont2
  A\kern-.1667em\lower.5ex\hbox{M}\kern-.125emS}}

\title{
{
\vspace{-3.0cm} \normalsize \hfill
\parbox{30mm}{COLO-HEP-475\\October 2001}
}\\[15mm]
       Simulating dynamical fermions with smeared links
       }

\author{A. Hasenfratz\address[MCSD]{Department of Physics, University of Colorado, 
        Campus Box 390, Boulder, CO 80309, USA}
        and
        F. Knechtli\addressmark[MCSD]
        \thanks{new address: Institut f{\"u}r Physik, 
        Humboldt-Universit{\"a}t zu Berlin, Invalidenstr. 110, 
        10115 Berlin, Germany}
       }

\begin{document}

\begin{abstract}
Smearing the gauge links of dynamical configurations removes small scale unphysical
vacuum fluctuations and thus improves the chiral properties of lattice fermions. 
Recently we proposed the hypercubic smearing (HYP) that improves the flavor symmetry
of staggered fermions by an order of magnitude with only minimal distortions at 
small distances. We describe a new algorithm to simulate dynamical HYP fermions
based on the standard pure gauge overrelaxation and heatbath updates. The algorithm
has been used to simulate four and two flavors of staggered fermions. Unlike
standard dynamical simulation techniques, this algorithm does not loose
efficiency at small quark masses.
\vspace{1pc}
\end{abstract}

\maketitle

\section{Introduction}

In this article we study an algorithm for simulating a fermionic
system described by the action
\begin{equation}\label{action}
S = S_g(U)-{\rm tr}\ln\left[Q^{\dagger}(V)Q(V)\right] \,,
\end{equation}
where $S_g(U)$ is the pure gauge action and $Q(V)$ is the fermionic matrix.
The gauge connections between fermions are smeared links $V$ which
are constructed deterministically from the dynamical thin links $U$. Since each
smeared link is a local combination of a finite number of thin links,
the system where the
fermions couple via smeared links is in the same universality class as the system
with thin links.\\
It has been demonstrated that smeared links improve flavor symmetry with staggered
fermions, chiral symmetry with Wilson-type clover fermions and they are useful
in the construction of an overlap Dirac operator. The problem is that the standard
simulation algorithms for dynamical fermions (Hybrid Monte Carlo (HMC)
or R algorithms)
involve the computation of the gauge force $\partial Q(V)/\partial U$, which is
either very complicated or likely impossible if the smeared links are
projected onto SU(3).


\section{New algorithm}

We propose for the simulation of the system described by eq. (\ref{action}) a
two step updating
\begin{itemize}
\item[1)] updating of a set of thin links $\{U\}\rightarrow\{U^{\prime}\}$ which
satisfies detailed balance for $S_g(U)$
\item[2)] accept/reject step with acceptance probability
\begin{equation}\label{accrej}
P_{\rm acc} = \min\left\{1,
  \frac{\det\left[Q^{\dagger}(V^{\prime})Q(V^{\prime})\right]}
       {\det\left[Q^{\dagger}(V)Q(V)\right]}\right\} \,,
\end{equation}
where $V^{\prime}$ are the smeared links constructed from the updated thin links
\end{itemize}
As concerns step 1) our choice for $S_g$ is the Wilson plaquette action and the
updating is performed either with microcanonical overrelaxation or
with Cabibbo-Marinari heatbath. With overrelaxation we update
all links within some finite block of the lattice (the sequence of the updates
in the block has to be symmetrized), with heatbath the links to be updated are
chosen randomly.\\
In step 2), we use a stochastic estimator to evaluate the ratio of fermionic
determinants
\begin{eqnarray}
P_{\rm acc}^{\prime} & = & \min\left\{1,\exp{\Delta S}\right\} \label{pacc1}
\\
\Delta S & = & 
\xi^{\dagger}[Q^{\dagger}(V^{\prime})Q(V^{\prime})-Q^{\dagger}(V)Q(V)]\xi \,,
\label{pacc2}
\end{eqnarray}
where the vector $\xi$ is generated according to the probability distribution
\begin{equation}\label{pacc3}
P(\xi) \propto \exp\{-\xi^{\dagger}Q^{\dagger}(V^{\prime})Q(V^{\prime})\xi\} \,.
\end{equation}

\section{The HYP action}

In the following we describe the application of this algorithm to the cases of
$n_f=4$ and $n_f=2$ flavors of staggered fermions coupled via hypercubic (HYP)
smeared links. The definition of HYP links and their properties are discussed in
\cite{coloquench,roland01}. In the natural formulation of four flavors of staggered
quarks, the continuum
SU(4) flavor symmetry is broken to U(1) at finite lattice spacing. The pion
spectrum has only one true Goldstone pion $\pi_{\rm G}$, the other 14 pions remain
massive when the quark mass goes to zero. The mass splitting between a
non-Goldstone pion $\pi$ and $\pi_{\rm G}$ can be parametrized by the quantity
\cite{milcdelta2}
\begin{equation}\label{delta2}
\delta_2 =
\frac{m_{\pi}^2-m_{\pi_{\rm G}}^2}{m_{\rho}^2-m_{\pi_{\rm G}}^2}
\end{equation}
evaluated at $m_{\pi_{\rm G}}/m_{\rho}=0.55$.
With a flavor symmetric action $\delta_2=0$ for all the pions $\pi$.
We computed $\delta_2$ for the two lightest non-Goldstone pions
(for the notation see \cite{coloquench}) on a set of quenched
$8^3\times24$ lattices generated with Wilson pure gauge action at $\beta=5.7$
($a=0.17\,{\rm fm}$),
using three different valence quark actions: standard staggered action with thin
and HYP smeared links and improved Asqtad action of the MILC collaboration
\cite{asqtad}. The results are shown in table \ref{tdelta2}. 
\begin{table}[h]
 \caption{Flavor symmetry breaking for different valence quark actions at
 lattice spacing $a=0.17\,{\rm fm}$. \label{tdelta2}}
 \begin{tabular}{|c|c|c|c|} \hline
  Action/$\delta_2$ & Thin & Asqtad & HYP
  \\ \hline\hline
  $\pi_{i,5}$ & 0.594(25) & 0.191(22) & 0.086(14) \\ \hline
  $\pi_{i,j}$ & 0.72(6) & 0.32(4) & 0.150(24) \\ \hline
 \end{tabular}
\end{table}
Flavor symmetry violations with the HYP action are reduced by an order of
magnitude with respect to the thin link action and by a factor of two with
respect to the Asqtad action.

\section{$n_f=4$ HYP staggered fermions}

We consider the matrix $Q$ in eq. (\ref{action}) given by
\begin{eqnarray}
Q_{i,j} & = & M_{i,j}\equiv
2(am)\delta_{i,j}+D_{i,j}(V) \,, \label{matrixnf4} \\
D_{i,j}(V) & = & 
\sum_{\mu}\eta_{i,\mu}(V_{i,\mu}\delta_{i,j-\hat{\mu}}-
 V^{\dagger}_{i-\hat{\mu},\mu}\delta_{i,j+\hat{\mu}}) \,, \nonumber
\end{eqnarray}
which describes four flavors of staggered fermions coupled via HYP
smeared links $V$. In the accept/reject step we rewrite the fermion matrix
as
\begin{eqnarray}
 Q(V) & = & Q_r(V)A(V) \,, \label{split} \\
 A(V) & = & \exp[\alpha_4D^4(V)+\alpha_2D^2(V)] \,. \label{amat}
\end{eqnarray}
In this way we achieve that an effective action
\begin{equation}
S_{\rm eff} =
-2\alpha_4\rm{Re}\,{\rm tr}D^4
-2\alpha_2\rm{Re}\,{\rm tr}D^2
\end{equation}
is removed from the fermion determinant. The acceptance probability 
in eq. (\ref{accrej}) becomes $\tilde{P}_{\rm acc}=$
\begin{equation}
\min\left\{1,
\exp[S_{\rm eff}(V)-S_{\rm eff}(V^{\prime})]\,\exp\Delta S_r\right\} \,,
\end{equation}
where the stochastic part $\Delta S_r$ is computed like in eqs.
(\ref{pacc2}-\ref{pacc3}) with $Q_r$ instead of $Q$. The real parameters
$\alpha_4$ and $\alpha_2$ are optimized to maximize $\tilde{P}_{\rm acc}$
and we use $\alpha_4=-0.006$, $\alpha_2=-0.18$ \cite{coloaux}. The point we
would like to emphasize is that this algorithm is not effective with thin links.
Smeared links constrain the fluctuations of the stochastic estimator 
$\exp\Delta S_r$ and make the algorithm efficient.\\
In Monte Carlo simulations we separate the measurements of the observables
by $N_{\rm OR}$ overrelaxation and $N_{\rm HB}$ heatbath two-step updatings,
changing $t_{\rm OR}$ and $t_{\rm HB}$ links respectively. At a lattice spacing
$a\sim0.17\,{\rm fm}$ and correlation length $m_{\pi_{\rm G}}r_0=2.0$
we simulated the HYP algorithm on a $8^3\times24$ lattice
with ($t_{\rm OR}=128$, $N_{\rm OR}=160$) and ($t_{\rm HB}=200$, $N_{\rm HB}=80$)
with an acceptance of $\sim20\%$. 
If we run the standard thin link action with HMC at approximately matched
physical parameters,
we obtain comparable autocorrelations in the observables when the
measurements are separated by a trajectory of unit time length.
We can then compare the time costs of the two algorithms which are
about 7 times larger for the HYP algorithm.\\
With the new algorithm the quark mass can be lowered to values impractical with
standard fermionic algorithms. A similar test as described above at
$m_{\pi_{\rm G}}r_0=1.6$ shows that the HYP algorithm is about 3 times slower
than the standard thin link HMC.\\
A drawback of our algorithm is that as the lattice volume increases the numbers
of the updated links $t_{\rm OR}$ and $t_{\rm HB}$ have to be kept unchanged,
consequently the numbers of updatings $N_{\rm OR}$ and $N_{\rm HB}$
have to be scaled with the volume to keep the autocorrelation times unchanged.
On the other hand as the continuum limit is approached the numbers of
links $t_{\rm OR}$ and $t_{\rm HB}$ which can be effectively updated scale: 
the physical volume of the updated region is constant.

\section{$n_f=2$ HYP staggered fermions}

We consider eq. (\ref{action}) with
\begin{equation}\label{defnf2}
Q^{\dagger}Q = (M^{\dagger}M)^{1/2} \,,
\end{equation}
where $M$ is the four flavor staggered fermion matrix.
Perturbative arguments indicate that
eq. (\ref{defnf2}) is equivalent in the continuum to a non-local theory of
two flavors of quarks \cite{sqroot}.\\
To evaluate the square root in eq. (\ref{defnf2}) we use a
polynomial approximation. We write $x^{1/2}=x x^{-1/2} = xP_{-1/2}(x)$ with
\begin{equation}\label{polyapprox}
P_{-1/2}(x) = \lim_{n\to\infty}P^{(n)}_{-1/2}(x) =
\lim_{n\to\infty}\sum_{i=0}^n c_i^{(n)} x^i \,.
\end{equation}
The coefficients $c_i$ are found by minimizing \cite{montvayp}
\begin{equation}\label{minipoly}
I = \int_0^{\lambda}dx\,(x^{-1/2}-P_{-1/2}^{(n)}(x))^2x \,.
\end{equation}
For $n$ even the roots $r_i$ of $P^{(n)}_{-1/2}$ come in complex conjugate pairs,
we can write
\begin{equation}\label{polyroots}
P_{-1/2}^{(n)}(x) = q_{(-1/2)}^{(n)}(x)q_{(-1/2)}^{(n)\dagger}(x) \,,
\end{equation}
where $q_{(-1/2)}^{(n)}(x)=\sqrt{c_n^{(n)}}\prod_{i=1}^{n/2}(x-r_i^{(n)})$.
This allows us to identify
\begin{equation}\label{matrixnf2}
Q = \lim_{n\to\infty}Mq_{(-1/2)}^{(n)}(M^{\dagger}M) \,.
\end{equation}
We note that the spectrum
of $M^{\dagger}M=4(am)^2-D^2$ is bound from below exactly by $4(am)^2$ and the
maximum eigenvalue fluctuates around the free field value $16+4(am)^2$. We
set $\lambda=1$ in eq. (\ref{minipoly}) and then rescale the polynomial to use
it in the range $x\in(0,18)$.\\
In the accept/reject step eq. (\ref{pacc1}) the stochastic estimator
$\exp\Delta S$ is evaluated as
\begin{eqnarray}
\Delta S^{(n)} & = &
\xi^{\dagger}[P_{-1/2}^{\prime(n)}M^{\prime\dagger}M^{\prime}
-P_{-1/2}^{(n)}M^{\dagger}M]\xi \label{nf2deltaS} \\
\xi & = & q_{-1/2}^{\prime(m)\dagger}R \,, \label{nf2xi}
\end{eqnarray}
where $M^{\prime}\equiv M(V^{\prime})$, etc. and $R$ is a random Gaussian vector.
The systematic error due to the finite order $n$ of the polynomial
is $\Delta^{(n)}=P_{-1/2}^{(n)}-P_{-1/2}$. It
is possible to improve eq. (\ref{nf2deltaS}) such that
\begin{equation}\label{deltaSimp}
(\Delta S^{(n)})_{\rm imp} = \Delta S^{(\infty)} + \mathcal{O}((\Delta^{(n)})^2) \,.
\end{equation}
A similar improvement is possible for the vector $\xi$ in eq. (\ref{nf2xi})
but we decided to take a higher order $m=128$ of the polynomial. In the actual
computation in order to increase the acceptance probability of the stochastic
estimator we split the fermion matrix $M=M_rA$ like in the $n_f=4$ case,
eqs. (\ref{split}-\ref{amat}).\\
In the evaluation of the polynomial eq. (\ref{polyroots}) the ordering of the
roots $r_i$ is important for the reduction of numerical round-off errors.
We emphasize
that the smeared links make the polynomial approximation work with lower order
polynomials.\\
In order to test the accuracy of the polynomial approximation
we looked at the quantities
\begin{equation}\label{deltapacc}
\Delta P_{\rm acc}^{(n,n^{\prime})} =
P_{\rm  acc}^{(n)}-P_{\rm acc}^{(n^{\prime})} \,,
\end{equation}
where $P_{\rm acc}^{(n)}=\min\{1,\exp(\Delta S^{(n)})_{\rm imp}\}$ is the
acceptance probability using $n$th order polynomial, comparing
$n=32,64,128$ with $n^{\prime}=256$. We found that
$\Delta P_{\rm acc}=\mathcal{O}(10^{-2})$ for $n=32$ and decreases to
$\mathcal{O}(10^{-4})$ for $n=128$. Moreover $\Delta P_{\rm acc}$ does not
increase significantly as the mass decreases or the lattice volume increases or
the lattice spacing decreases. To answer the question of which accuracy is
needed, we performed simulations with different values $n=32,64,128$ but all
other parameters fixed. First numerical results for the chiral condensate show
agreement within the statistical errors for all orders $n$ of the polynomial
but much longer runs are needed to identify the systematic errors of a low order
polynomial. The properties of the polynomial approximation have been studied
extensively in \cite{montvayp} and we plan to follow some of those methods in
the future.

\end{document}